\documentclass[prl,twocolumn,showpacs,floatfix,amsfonts]{revtex4}
\usepackage{graphicx,graphics,color,epsfig}
\usepackage{bm}
\usepackage{amsmath}
\usepackage{amssymb}
 
\def\Im{\text{Im}} 
\def\sgn{\text{sgn}}

\def\be{\begin{equation}} 
\def\ee{\end{equation}} 
\def\bea{\begin{eqnarray}} 
\def\eea{\end{eqnarray}}
\begin{document}
\preprint{}
\title{Quantum critical properties of the Bose-Fermi Kondo model
in a large-$N$ limit}
\author{Lijun Zhu$^{a}$, 
Stefan Kirchner$^{a}$, 
Qimiao Si$^{a}$, 
and Antoine Georges$^{b}$}
\affiliation
{$^{a}$Department of Physics \& Astronomy, Rice University, Houston, 
TX 77005-1892, USA \\
$^{b}$Centre de Physique Th{\'e}orique, Ecole Polytechnique, 91128 
Palaiseau Cedex, France}


\begin{abstract}
Studies of 
non-Fermi liquid properties in heavy fermions
have led to the current interest in the Bose-Fermi Kondo model.
Here we 
use a dynamical large-$N$ 
approach to analyze an
SU($N$)$\times$SU($\kappa N$) generalization of the model.
We establish the existence
in this limit
of an unstable fixed point  
when the bosonic bath has a sub-ohmic
spectrum ($|\omega|^{1-\epsilon} \sgn \omega$, with $0<\epsilon<1$).
At the quantum critical point, 
the Kondo scale vanishes and the local spin susceptibility
(which is finite on the Kondo side for $\kappa <1$)
diverges.
We also find an $\omega/T$ scaling
for an extended range (15 decades) of $\omega/T$.
This scaling 
violates (for $\epsilon \ge 1/2$)
the expectation of a naive mapping to certain classical
models in an extra dimension; it reflects the inherent
quantum nature of the critical point.
\end{abstract}
\maketitle


Heavy fermion systems show unusual quantum critical properties,
which have motivated the interest in how the Kondo effect
is destroyed near a magnetic quantum phase 
transition\cite{Sietal2001,Colemanetal2001,Burdinetal2002,Senthiletal2003}.
One theoretical approach studies Kondo lattice systems through 
a self-consistent Bose-Fermi Kondo model\cite{Sietal2001}.
Here, a local magnetic moment not only 
interacts antiferromagnetically with the spins of a conduction electron
bath, it also is coupled to a dissipative bosonic bath.
(For work in a related context, see Ref.~\onlinecite{SenguptaGeorges}.)
The conduction electrons alone would lead to a Kondo
singlet for the ground state and 
spin-$\frac{1}{2}$ charge-$e$
Kondo resonances in the excitation spectrum\cite{Hewson}. 
The bosonic bath -- 
characterizing the fluctuating magnetic field
provided by spin fluctuations -- competes with the Kondo effect;
a sufficiently strong bosonic coupling destroys the Kondo effect.
A quantum critical point (QCP) arises, where 
the electronic excitations are of non-Fermi liquid nature. 
When such a Kondo-destroying criticality
is embedded into the criticality
of a magnetic ordering,
the critical theory becomes distinct from the standard
paramagnon theory\cite{Sietal2001}.
This picture has been called locally quantum critical.
Direct evidence for the destruction of the Kondo effect has 
recently emerged in the Hall measurements of a
heavy fermion metal\cite{Paschen}.

Previous studies of the Bose-Fermi Kondo model
are based on an $\epsilon$-expansion renormalization group (RG)
analysis\cite{SmithSi,Sengupta,Sietal2001,Vojta1,Zhu-Si,Zarand}
[the definition of $\epsilon$ appears in Eq.~(\ref{Aphi}) below;
$\epsilon>0$ corresponds to sub-ohmic bosonic bath].
This approach has been successful, partly due to the
fact that the linear in $\epsilon$ contribution 
to the anomalous dimension of the local
spin operator turns out to be exact to all orders 
in $\epsilon$~\cite{Vojta1,Zhu-Si,Zarand}.
In spite of this, the approach is after all perturbative in
$\epsilon$; moreover, it is capable of calculating only
a limited number of physical quantities.
(The model lacks conformal invariance,
making boundary conformal field theory inapplicable.)
For the case with an Ising spin-anisotropy, a self-consistent 
version of the model 
has been studied\cite{GrempelSi}
by a Quantum Monte-Carlo method\cite{GrempelRozenberg}.
With spin rotational invariance, the only approach
that has been used, beyond the $\epsilon$-expansion,
is the condensed slave-boson mean field theory\cite{Burdin};
here,
the QCP is trivial: 
the effect of the Kondo 
interaction disappears completely as soon as the static
slave-boson amplitude vanishes.
A dynamical method is necessary 
to capture the
quantum critical properties.

In this paper,
we study the 
model using  
a dynamical large-$N$ method\cite{Parcollet1,Cox,SachdevYe}.
The spin symmetry is taken to be SU($N$) and the 
number of conduction electron channels $M \equiv \kappa N$. 
The large-$N$ limit is expected to capture the
quantum-critical behavior of the physical cases (see below).
This limit, with purely Kondo coupling to fermions was 
extensively studied in Refs.~\cite{Parcollet1,Cox}
and with purely bosonic coupling in Ref.~\cite{SachdevYe}
(see also \cite{Vojta1,Sengupta}).

Our primary results are three-fold. First, we establish
the existence of a non-trivial 
QCP 
even in the large-$N$ limit.
Second, 
we show that the QCP is non-Gaussian not only for
small $\epsilon$ but also for $\epsilon \ge 1/2$.
This demonstrates the inherently quantum nature of the QCP:
standard mapping of a quantum critical point to a classical critical
point at extra dimensions\cite{Sachdev-book} would imply that
$\epsilon=1/2$ is the upper critical dimension
({\it c.f.} the classical spin chains with $1/r^{2-\epsilon}$ 
interactions\cite{Fisher,Kosterlitz});
the difference arises from the Berry phase of the quantum spin.
Third, we determine the critical exponents
and amplitudes and the universal scaling functions 
near the QCP,
as well as
the properties on the approach towards the bosonic fixed point.
We will restrict to $\kappa \equiv M/N <1$ for which
the spin susceptibility of the multi-channel
Kondo fixed point is finite\cite{Parcollet1,Cox,Affleck-Ludwig}.
 
The Hamiltonian for the 
model is
\begin{eqnarray} 
{\cal H}_{\text{MBFK}} &=& 
({J_K}/{N}) 
\sum_{\alpha}{\bf S} 
\cdot {\bf s}_{\alpha} 
+ \sum_{p,\alpha,\sigma} E_{p}~c_{p \alpha 
\sigma}^{\dagger} c_{p \alpha \sigma} 
\nonumber\\ 
&+& 
({g}/{\sqrt{N}})
{\bf S} \cdot 
{\bf \Phi}
+ \sum_{p} 
w_{p}\,{\bf \Phi}_{p}^{\;\dagger}\cdot {\bf \Phi}_{p}. 
\label{H-MBFK} 
\end{eqnarray} 
Here, a local moment ${\bf S}$ interacts with 
fermions
$c_{p\alpha\sigma}$ {\it and} 
bosons ${\bf \Phi}_p$. 
The spin and channel indices are $\sigma = 1, \ldots, N$
and 
$\alpha=1, \ldots, M$,
respectively,
and ${\bf \Phi} \equiv \sum_p ( {\bf \Phi}_{p} + 
{\bf \Phi}_{-p}^{\;\dagger} )$ contains $N^2-1$
components. We will first consider a flat 
conduction electron density of states
and a subohmic bosonic spectrum.
Denoting ${\cal G}_0 = - \langle
T_{\tau} c_{\sigma\alpha}(\tau) c_{\sigma\alpha}^{\dagger}(0)
\rangle _0$ and 
${\cal G}_{\Phi} =  \langle
T_{\tau} \Phi(\tau) \Phi^{\dagger}(0)
\rangle _0$ for each component, we have
$N_0(\omega) \equiv - {1\over \pi} \Im {\cal G}_{0}(\omega+i0^+) 
=N_0$
($|\omega|< D = 1/2N_0$)
and $A_{\Phi}(\omega) \equiv {1\over \pi} \Im {\cal G}_{\Phi}(\omega+i0^+) 
=\sum_p[\delta (\omega - w_p) -\delta (\omega + w_p)]$
being
\bea
A_{\Phi}(\omega) = 
[K_0^2 / \Gamma(2-\epsilon)] |\omega|^{1-\epsilon} \sgn(\omega) ,
\label{Aphi}
\eea
for $|\omega| 
<
\Lambda \equiv 1/\tau_0$
($\Gamma$ is the gamma function).
We consider 
the conduction electrons
in the fundamental representation of the SU($N$)$\times$SU($M$)
group, and the local moment in an antisymmetric representation
whose Young tableaux is a single column of $N/2$ boxes. 
We can then write $S$
in terms of pseudo-fermions,
$S_{\sigma 
\sigma'}=f_{\sigma}^{\dag}f_{\sigma'}-
\delta_{\sigma, \sigma'}/2$, 
with the constraint $\sum_{\sigma=1}^N 
f_{\sigma}^{\dag}f_{\sigma}= 
N/2$, which is
enforced by a Lagrange multiplier $i\mu$. 
The Kondo coupling,
$(J_{K}/N)\sum_{\sigma \sigma'} \left( 
f_{\sigma}^{\dag}f_{\sigma'}- 
\delta_{\sigma, \sigma'} /2
\right) c_{\alpha\sigma'}^{\dagger}c_{\alpha\sigma} $,
for each channel $\alpha$, will be decoupled into
a $B_{\alpha}^{\dag}\sum_{\sigma}c_{\alpha 
\sigma}^{\dag}f_{\sigma}/\sqrt{N}$ interaction using a dynamical
Hubbard-Stratonovich field $B_{\alpha}(\tau)$.
It is convenient to consider $g$ and $J_K$ as independently
varying, since the universal properties only depend
on the ratio $g/T_K^0$.

{\it Renormalization group analysis:~} 
We will first establish that 
a non-trivial QCP survives the 
large-$N$ limit. We have generalized the RG equations
of Ref.~\onlinecite{Zhu-Si} to arbitrary $N$ and $M$. In the 
large-$N$ limit, the RG beta functions, perturbative only 
in $\epsilon$, become
\begin{eqnarray} 
\beta(\bar{g}) =&& -2 \bar{g}\left( 
{\epsilon}/{2}
-\bar{g} + 
\bar{g}^2 
    -\kappa \bar{j}^2\right),  \nonumber \\ 
\beta(\bar{j}) =&&  -\bar{j} \left( \bar{j} - \kappa \bar{j}^2 
  -\bar{g} + \bar{g}^2 \right), 
\label{rg-equations} 
\end{eqnarray} 
where, $\bar{g} \equiv (K_0 g)^2$, and $\bar{j} \equiv N_0 J_K$. 
There is an unstable fixed point at
$(\bar{g}^*,\bar{j}^*) = 
({\epsilon}/2 + (1-\kappa)\epsilon^2/4, 
{\epsilon}/2)$.
Here, 
$\chi (\tau) \sim 
{1}/{|\tau|^{\eta}}$
and the anomalous dimension
$\eta = \epsilon$.

\begin{figure} 
\includegraphics[width=0.3\textwidth]{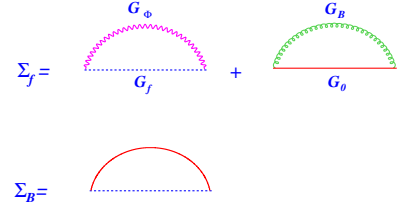} 
\caption{The large-$N$ Feynman diagrams for the self-energies.}
\label{fg-diag} 
\end{figure}

{\it Saddle-point equations:~}
Our primary objective
is to study the saddle-point
equations of the large-$N$ limit,
which have the following form
({\it cf.} Fig.~\ref{fg-diag}):
\begin{eqnarray} 
G_B^{-1}( i\omega_n) &=& 
1/{J_K} 
- \Sigma_B( i\omega_n);
~~~
\Sigma_B(\tau) = - {\cal G}_{0}(\tau) G_f(-\tau)%
 \nonumber \\
G_f^{-1}(i\omega_n)& =& i\omega_n - \lambda -
\Sigma_f(i\omega_n) ; \nonumber \\ 
\Sigma_f(\tau)&=& \kappa {\cal G}_{0}(\tau) G_B(\tau) + g^2 
G_f(\tau){\cal G}_{\Phi}(\tau) ,
\label{NCA}
\end{eqnarray} 
together with a constraint equation,
\begin{eqnarray} 
G_f(\tau = 0^{-}) = ({1}/{\beta}) \sum_{i\omega_n}
G_f (i\omega_n) {\rm e}^{i\omega_n 0^+} = {1}/{2} .
\label{constraint}
\end{eqnarray} 
The dynamical spin susceptibility
is 
\begin{eqnarray} 
\chi (\tau) = - G_f (\tau) G_f (-\tau) .
\label{chi-tau}
\end{eqnarray} 
Here, $\lambda$ is the saddle point value of $i \mu$,
$G_{B}(\tau) = \langle
T_{\tau} B_{\alpha} (\tau) 
B_{\alpha}^{\dagger} (0)
\rangle $ and $G_{f}(\tau) =
- \langle
T_{\tau} f_{\sigma} (\tau) f_{\sigma}^{\dagger}(0)
\rangle $.

The first and second terms on the RHS of 
the last line of Eq.~(\ref{NCA}) reflect 
the Kondo and bosonic couplings, respectively. They capture the competition
between the two types of interactions. 
To proceed, we seek for scale-invariant solutions
of the following form at $T=0$,
\begin{eqnarray} 
G_f(\tau)&=& {A_1}/{|\tau|^{\alpha_1}} 
\sgn \tau ,
~~~~~~G_B(\tau) = 
{B_1}/{|\tau|^{\beta_1}} ,
\end{eqnarray} 
for the asymptotic regime $\tau \gg \tau_0$.
Using these,
the saddle-point equations become
\begin{eqnarray} 
G_f^{-1}(\omega) &=&
{ \omega^ {1-\alpha_1} \over A_1 C_{\alpha_1-1}} 
= \omega
+ \kappa N_0 B_1 C_{\beta_1} 
\omega^{\beta_1} 
\nonumber \\ 
&& - (K_0 g)^2 A_1 C_{1-\epsilon+\alpha_1} 
\omega^{1-\epsilon+\alpha_1} 
\label{saddle.gf}
\\
G_B^{-1}(\omega) &=& 
{ \omega^ {1-\beta_1} \over 
{B_1 
\beta_1C_{\beta_1}}} 
\nonumber\\
&=&
{1 \over J_K} - \Sigma_B(0) + N_0 A_1 
(\alpha_1+1) C_{\alpha_1+1} 
\omega^{\alpha_1} 
\label{saddle.gb}
\end{eqnarray} 
where $C_{x-1} \equiv \frac{\pi}{\Gamma(x)}
\frac{\exp [i\pi (2-x)/2]}{\sin [\pi (2-x)/2]}$.
We have used $\lambda +\Sigma_f(0)=0$, which follows from
Eq.~(\ref{constraint}).
We have also assumed $0<\alpha_1,\beta_1 < 1$, which turns out to
be valid in most cases; exceptions will be 
specified below.

Three solutions arise depending on the competition between the
Kondo and bosonic coupling terms.
A dominating Kondo or bosonic term leads to the Kondo 
or bosonic phase,
respectively. When the two terms are 
of the same order, the critical fixed point emerges.

{\it Multichannel Kondo and bosonic fixed points:~} 
When $\beta_1 < 1-\epsilon+\alpha_1$,
the Kondo coupling dominates over the bosonic coupling
on the RHS of Eq.~(\ref{saddle.gf}).
The leading order solution is then identical 
to that of the model with only a Kondo coupling\cite{Parcollet1}.
The leading exponents are 
$\alpha_1 = 1/(1+\kappa)$ and $\beta_1 = 
\kappa /(1+\kappa )$.
 
In the opposite case, with $\beta_1 > 1-\epsilon+\alpha_1$,
the bosonic term dominates over the Kondo term.
Matching the dynamical parts of 
Eq.~(\ref{saddle.gf})
leads to
\bea 
\alpha_1&=&\epsilon/2 
; ~~~
A_1^2 = 
{{2-\epsilon} \over {4 \pi
(K_0g)^2 }} 
\tan{\pi \epsilon \over 4}. 
\eea 
Together with Eq.~(\ref{chi-tau}),
they combine to yield the following result for the 
dynamical spin susceptibility
\bea 
\chi(\tau)=  \left({2-\epsilon \over 4 \pi} 
\tan{\pi \epsilon \over 4}\right) \frac{1}{(K_0 g)^2}
{1 \over |\tau|^{\epsilon}} .
\label{chi-bosonic}
\eea 
The exponent agrees with that of
Refs.~\onlinecite{SachdevYe,SmithSi,Sengupta,Vojta1,Zhu-Si,Zarand}.

If $J_K$ is strictly $0$, $G_B(\omega)$ vanishes. For small but finite
$J_K$, Eq.~(\ref{saddle.gb}), which (more precisely, the term in between
the two equalities) comes with the assumption $\beta_1<1$, cannot 
be satisfied. A solution does however exist for $\beta_1>1$, in which
case $G_B(\omega)$ no longer diverges.
It now follows from $G_B^{-1}(\omega)
\sim \text{const} + \omega^{\beta_1 -1 } $ that
\bea 
\beta_1=1+\epsilon/2 .
\label{beta1-bosonic}
\eea 
The non-divergence of $G_B$ reflects the irrelevant nature 
of the Kondo interaction at the bosonic fixed point.

{\it Critical fixed point:~} 
The Kondo and bosonic terms are of the same order when
$\beta_1 = 1-\epsilon+\alpha_1$. In this case, 
matching the dynamical parts leads to
\bea
\alpha_1 = \epsilon/2
; ~~~~~~
\beta_1 = 1- \epsilon/2; ~~~~~~~~~~~\\
A_1^2 = 
{2-(1+\kappa )\epsilon \over 
{4 \pi (K_0g_c)^2 } }
\tan{\pi \epsilon \over 4} 
;
~
B_1 = 
- {\epsilon \over 4 \pi N_0A_1} \tan{\pi \epsilon \over 4} .
\eea 
The critical bosonic coupling $g_c$
needs to be numerically determined by matching 
the static part:
$1/J_K - \Sigma_B(0)=0$.
The resulting dynamical spin susceptibility is:
\bea 
\chi(\tau)= \left( {{2-(1+\kappa)\epsilon} \over {4 \pi} }
\tan{\pi \epsilon \over 4} \right) \frac{1}{(K_0 g_c)^2}
{1 \over |\tau|^{\epsilon}} .
\label{chi-critical}
\eea 

We make two observations at this point. First, for 
infinitesimal $\epsilon$,
our result agrees not only with the result of the $\epsilon$-expansion
for the large-N model [{\it c.f.} the subsection containing 
Eq.~(\ref{rg-equations})] but also that 
of the $\epsilon$-expansion for the the physically relevant
SU(2) single-channel model\cite{Zhu-Si,Zarand}. This implies
that the large-N limit captures the quantum critical behavior
of the physical systems, even though it yields a multichannel
behavior on the Kondo side. This is reminiscent of the effects
of spin anisotropies in the single-channel models: though
the bosonic fixed point with Ising anisotropy (whose local
susceptibility contains a finite Curie constant\cite{Zhu-Si})
is very different from its counterpart with SU(2) symmetry,
$\eta$ is the same for the QCPs of the two cases\cite{Zhu-Si}.
Second, Eq.~(\ref{chi-critical}) goes beyond the $\epsilon$-expansion
result described earlier in the sense that it is valid
beyond infinitesimal $\epsilon$.

We have also determined the exponents of the sub-leading
terms for $G_f$ and $G_B$, finding 
$\alpha_2= \epsilon$ and $\beta_2 = 1$.

\begin{figure} 
\includegraphics[width=0.5\textwidth]{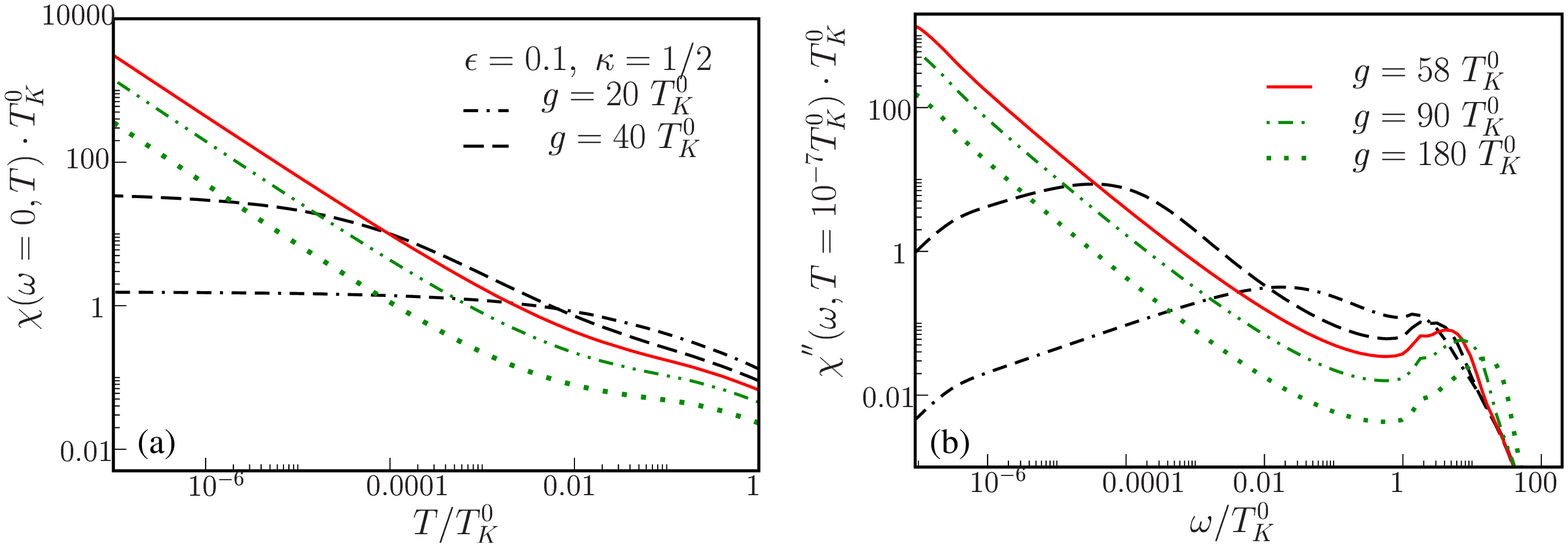} 
\caption{Static and dynamical local susceptibilities for $\epsilon=0.1$.
The red solid line corresponds to the 
quantum critical point, $g=g_c$.}
\label{suscep-epsilon=0.1}
\end{figure} 

We next consider approaching the QCP from the Kondo side.
Defining $T^*$ to be the crossover scale where the spectral function
$A_f(\omega)$ changes from 
$A_f(\omega) \sim \frac{1}{\omega^{\kappa/(1+\kappa)}}$ 
for $|\omega| \ll T^*$
to $A_f(\omega) \sim 
\frac{1}{\omega^{1-\epsilon/2}}$ for $|\omega| \gg T^*$,
and inserting the ansatz into $1/J_K = \Sigma_B(\omega=0)
=\int_0^{\beta} d \tau {\cal G}_0(\tau)
G_f(\tau)$,
we find
$T^* \sim (g_c-g)^{2/\epsilon}$.
Correspondingly, the susceptibility diverges as $g$ approaches
$g_c$:
\bea
\chi (T,\omega=0,g \alt g_c) \sim (g_c-g)^{-\gamma} ,~~
\gamma = 2(1 - \epsilon)/{\epsilon} .
\label{exponent-gamma}
\eea

Our results so far 
represent an essentially complete analytical solution at zero
temperature. The solution at finite temperatures is considerably more 
difficult to obtain.
In the following, we resort to numerical
studies.

{\it Numerical results:~} 
For low-energy analysis,
we find it important to
solve the integral eqs.
(\ref{saddle.gf},\ref{saddle.gb})
in {\it real frequencies}
and using a {\it logarithmic discretization}
[supplemented by 
linearly-spaced points;
typically 250 points in total for the frequency range
$(-10,10)$].
The convergence
is determined in terms of 
$A_B(\omega)$ and $A_f(\omega)$:
for each, the sum over the entire frequency
range of the relative difference between 
two consecutive iterations 
is less than $10^{-5}$. 
We
choose
$\kappa =1/2$
and
$N_0(\omega) = (1/\pi){\rm exp}(-\omega^2/\pi)$.
The nominal bare Kondo scale 
is
$T_K^0 \equiv (1/N_0) {\rm exp}(-1/N_0J_K) \approx 0.06$,
for fixed $J_K=0.8$.
The bosonic bath spectral function 
[{\it cf.} Eq.~(\ref{Aphi})]
is cut off smoothly
at $\Lambda \approx 0.05$.
$K_0^2$ is fixed at $\Gamma(2-\epsilon)$.

\begin{figure} 
\includegraphics[width=0.5\textwidth]{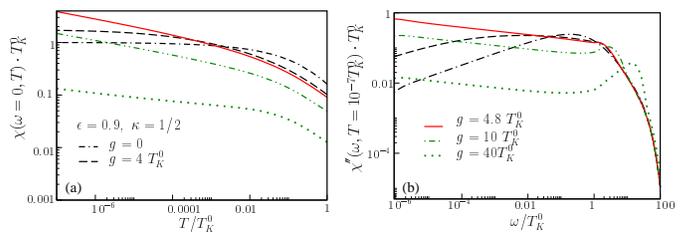} 
\caption{Static and dynamical susceptibilities for $\epsilon=0.9$.}
\label{suscep-epsilon=0.9}
\end{figure} 

In Fig.~\ref{suscep-epsilon=0.1}a), we show the static local spin
susceptibility $\chi(T)$ as a function of temperature
for $\epsilon=0.1$. For vanishing and relatively small 
bosonic coupling $g$,
$\chi(T)$ saturates to a finite value in the zero-temperature
limit -- which characterizes the Kondo phase. At a critical 
coupling\cite{footnote}
of $g/T_K^0 \approx 56$, $\chi(T)$ 
diverges in a power-law fashion.
Beyond this threshold coupling, 
$\chi(T)$ continues to have a power-law divergence.
Its amplitude, on the other hand, decreases with increasing $g$. 
These features are not restricted to small $\epsilon$;
similar results are given in
Fig.~3a), for $\epsilon=0.9$.
They are consistent with the analytical results:
Eq.~(\ref{exponent-gamma}) 
states
that
$\chi(T=0)$ should diverge as $g$ approaches $g_c$ from below;
for $g>g_c$, on the other hand, 
Eq.~(\ref{chi-bosonic}) implies that
the susceptibility amplitude should indeed
decrease as 
$g$ increases.

The quantum phase transition can also be seen in the dynamics.
In Fig.~2b),
we show the imaginary part of the dynamical local spin 
susceptibility, $\chi''(\omega)$, at the lowest studied
temperature ($T=10^{-7}T_K^0$) 
and for $\epsilon=0.1$. On the Kondo side, it vanishes
in the zero-frequency limit with an exponent close to $0.33$,
consistent with the analytical result $2\alpha_1-1 = 1/3$.
At $g_c$, it displays a power-law divergence, with an exponent close to
$0.9$. On the bosonic side, the exponent remains the same while
the amplitude decreases as $g$ increases. Both exponents agree with
the analytical result of $1-2\alpha_1=1-\epsilon=0.9$.
Similar results for
$\epsilon=0.9$ are shown in Fig.~3b).

\begin{figure} 
\includegraphics[width=0.45\textwidth]{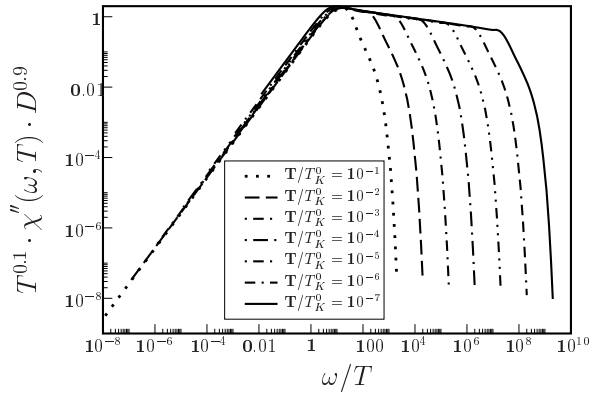} 
\caption{$\omega/T$ scaling of the 
dynamical spin susceptibility at $g=g_c$, 
for $\epsilon=0.9$ and $\kappa=1/2$.
The scaling exponent is $0.1$}
\label{fig-omega-T-scaling}
\end{figure}

We now 
address the 
finite frequency and temperature behavior at the QCP
in some detail.
Fig.~\ref{fig-omega-T-scaling} plots the dynamical
spin susceptibility for $\epsilon=0.9$, clearly manifesting 
an $\omega /T$ scaling. Remarkably, the scaling
covers an overall 15 decades of 
$\omega/T$ (from $10^{-8}$ to about $10^7$)! For each temperature,
the result falls on the scaling curve until an $\omega$
of the order $T_K^0$. 
Similar behavior is observed for 
other values of $\epsilon$ in the range
$0<\epsilon<1$.
The $\omega/T$ scaling
reflects
the interacting nature of the critical 
fixed point\cite{Sachdev-book}: 
there is no energy scale other than $T$; the relaxation rate,
originating from some relevant coupling, has to be linear in $T$.

The small but visible deviation from a complete collapse
in the range $\omega,T < T_K^0$ 
reflects the influence of subleading contributions.
For $\epsilon=0.9$, a similar degree of 
$\omega/T$ scaling is also seen in the $B$- and $f-$spectral
functions $A_B(\omega,T)$ and $A_f(\omega,T)$. 
The imaginary part of the $f$-self-energy, $\Sigma_f''(\omega,T)$,
on the other hand, starts to deviate from $\omega/T$ scaling at a
smaller frequency, reflecting a stronger effect of the subleading
terms. For $\epsilon=0.1$, $A_B(\omega,T)$ and 
$\Sigma_f''(\omega,T)$ are less influenced by the sub-leading 
terms than $A_f(\omega,T)$ and $\chi''(\omega,T)$.

Our work serves as a basis for 
studies of the spin-isotropic
Kondo lattice systems: by establishing the first non-perturbative
approach to the Kondo-destroying QCP of the Bose-Fermi Kondo model,
we are in the position to study the lattice problem in as systematic
a way as 
the Quantum Monte-Carlo study of the Kondo lattice 
with Ising anisotropy\cite{GrempelSi}.
Our results are also relevant in two other contexts.
First, some generalized version of the Bose-Fermi Kondo
model may capture the physics of impurities in high 
${\rm T_c}$
superconductors\cite{Vojta-Kircan,Bobroff}. Results similar to what
we report here should shed light on the dynamical properties 
of such systems, which have not yet been systematically addressed.
Second, 
the competition between Kondo and paramagnon couplings 
also appears
in the context of disorder in nearly magnetic 
metals\cite{Maebashi}.

We 
acknowledge the support of NSF Grant 
DMR-0424125 and the 
Robert A. 
Welch Foundation
(LZ, SK, and QS), DFG (SK), CNRS and Ecole Polytechnique
(AG) and NSF Grant 
PHY99-07949 (LZ, QS, AG), 
as well as the Lorentz Center, Leiden and 
KITP, UCSB.


\begin{thebibliography}{99}
\bibitem{footnote}
The dimensionless $g_c\Lambda^{1-\epsilon/2}/T_K^0$
is of order unity.
\end{thebibliography}
\end{document}